\documentclass[showpacs,preprintnumbers,amsmath,amssymb,twocolumn]{revtex4}
\usepackage{epsfig}
\usepackage{graphicx}
\usepackage{dcolumn}
\usepackage{bm}
\newcommand{\be}{\begin{equation}}
\newcommand{\ee}{\end{equation}}
\newcommand{\ba}{\begin{eqnarray}}
\newcommand{\ea}{\end{eqnarray}}

\begin{document}


\title{Is the anomalous decay ratio of $D_{sJ}(2632)$ due to isospin breaking? }
\author{L. Maiani}
\email{luciano.maiani@roma1.infn.it}
\affiliation{Universit\`{a} di Roma `La Sapienza' and I.N.F.N., Roma, Italy}
\author{F. Piccinini}
\email{fulvio.piccinini@pv.infn.it}
\affiliation{I.N.F.N. Sezione di Pavia and Dipartimento di Fisica Nucleare 
e Teorica, via A.~Bassi, 6, I-27100, Pavia, Italy}
\author{A.D. Polosa}
\email{antonio.polosa@cern.ch}
\affiliation{Centro Studi e Ricerche ``E. Fermi'', via Panisperna 89/A-00184
Roma, Italy}
\author{V. Riquer}
\email{veronica.riquer@cern.ch}
\affiliation{CERN Theory Department, CH-1211, Switzerland}


\begin{abstract}
Quark pair annihilation into gluons is suppressed 
at large momenta due to the asymptotic freedom. 
As a consequence, mass eigenvalues of heavy 
states should be almost diagonal with respect to  
up and down quark masses, thereby breaking isospin. 
We suggest the particle observed by the SELEX 
Collaboration, $D_{sJ}(2632)$ to be to a good extent a 
$[cd][{\bar d}{\bar s}]$ state, which would explain 
why its $D^0 K^+$ mode is anomalously suppressed with
respect to $D_s \eta$. Predictions for the rates of
the yet unobserved modes $D_s \pi^0$ and $D^+ K^0$ are given.
\newline\newline
ROME1-1381/2004, FNT/T-2004/11, BA-TH/488/04, CERN-PH-TH/2004-125

\end{abstract}

\maketitle

The SELEX Collaboration~\cite{fermilab} at Fermilab 
claims a $7\sigma$ observation of a narrow charmed meson, $D_{sJ}(2632)$, 
decaying into $D_s^+\eta$ and $D^0K^+$. The ratio between 
the $D^0K^+$ and the $D_s^+\eta$ modes reported is about $0.16\pm 0.06$. 
As the Collaboration points out, this is quite an anomalous result, 
given also that the decay momentum of the first mode is about twice that 
of the second. This  result would be totally at variance with the
attribution of the $D_{sJ}(2632)$ to a $c\bar s$ state. 
Pending confirmation of this effect and a determination of the particle 
quantum numbers, we point out in this note that this result 
would arise quite naturally if the $D_{sJ}(2632)$ were a bound state of a 
diquark-antidiquark pair, in particular an S-wave scalar~\cite{us}. 
The suppression of quark pair annihilation into gluons, due the asymptotic freedom, 
makes so that the mass eigenvalues are aligned with 
states diagonal with respect to quark masses, even for the light, 
up and down, quarks. The possibility of such an effect for pentaquark
states was pointed out in ref.~\cite{rossi-veneziano}. In our case it
is supported by the close degeneracy of a(980) and f(980) mesons, 
which should become more pronounced for the analogous states
at the charm energy scale.  
The $D_{sJ}(2632)$ would be essentially a 
$[cd][\bar d \bar s]$ state (not an isospin eigenstate) 
whose decay into $D^0K^+$ is forbidden by 
the Okubo-Zweig-Iizuka et al. rule~\cite{ozi}.
The interpretation proposed here is vulnerable to very simple tests, 
which we hope may be performed in the near future:\\1. The same $D_{sJ}(2632)$ 
resonance should decay into $D_s^+\pi^0$ and $D^+K^0$, with 
sizeable branching ratios which we predict within narrow bounds. 
Simultaneous decay into in $D_s^+\eta^0$ 
and $D_s^+\pi^0$ is direct proof of isospin breaking,\\2. 
A charge +2 state, very close in mass, should exist and be 
produced with sizeable cross section, mostly decaying 
into $D_s^+\pi^+$ and $D^+K^+$.

In~\cite{us} we propose 
that scalar mesons below 1 GeV are four quark states of the 
form $[qq][\bar q \bar q]$ ($q$= up, down, strange) where brackets 
represent states which are completely antisymmetric in color, flavor 
and spin. We show that this interpretation gives a good explanation 
of the spectrum and decay modes, except for the OZI 
rule violating mode $f\rightarrow \pi \pi$, which turns out to be 
larger than predicted. Decays are computed in terms of a single 
coupling, representing the amplitude for the switch of a $q \bar q$ 
pair between the diquarks, transforming the state into a pair of 
colorless mesons. A firm prediction of the scheme is the existence 
of similar scalar 
mesons with one light quark replaced by a heavy quark, e.g. charm. 
As discussed in ~\cite{us} we expect such particles to occur in a 
reducible ${\bf 6}\oplus\bar{\bf 3}$ of flavor $SU(3)$.
States with $C=S=+1$, of the 
form $[cq][\bar s \bar q]$ ($q$ is now restricted to up and down quarks)
form  an $I=1$ and $I=0$ complex of four states with electric 
charges 0, +1, +2. There are two states with electric charge~$+1$: 
$I=1$, $I_3=0$ and $I=0$. By analogy with the light scalar
meson complex, $a(980)$ and $f(980)$, we call the two states
$a^+_{c \bar s}$ and $f^+_{c \bar s}$.\\
If isospin were strictly conserved, the two states would be pure 
mass eigenstates belonging, respectively, to the ${\bf 6}$ and 
to the $\bar{\bf 3}$ and different decay modes. 
One expects~\cite{us} the four decay channels:
\begin{eqnarray}
a^+_{c \bar s}&=& \frac{([cu][\bar u \bar s]-
[cd] [\bar d \bar s])}{\sqrt{2}} \rightarrow D_s \pi^0,\;
(D K)_{\tiny{\begin{array}{c}
I=1 \\ I_3=0
\end{array}}},\nonumber\\ \nonumber
f^+_{c \bar s} &=& 
\frac{([cu][\bar u \bar s]+[cd] [\bar d \bar s])}{\sqrt{2}} 
\rightarrow D_s \eta,\;  (D K)_{I=0}.
\label{eq:a+f+}
\end{eqnarray}

The mesons $a(980)$ and $f(980)$ are degenerate within, say, 10~MeV~\cite{pdg}. 
As seen in~\cite{us}, this reflects the smallness of the OZI violating contributions 
to the mass matrix, which would align the mass eigenstates to pure $SU(3)$ 
representations. We expect OZI violations to be even smaller in heavy meson systems 
(as exemplified by the narrow width of the $J/\Psi$) and mass 
eigenstates to align strictly on the quark composition rather than the $SU(3)$, 
or even $SU(2)$ representations. This happens when the diagonal masses of 
the I=1 and I=0 states become degenerate within few MeV, comparable 
to the non-diagonal matrix element induced by the up and down quark 
mass difference (in a different context, a second order weak interaction is sufficient to 
maximally mix $K^0$ and $\bar{K^0}$, due the degeneracy of the 
diagonal masses implied by CPT). The issue of $SU(2)$ violation 
in mass eigenstates has been analyzed recently in ref.~\cite{rossi-veneziano} 
with the conclusion that considerable mixing between $I= 3/2$ 
and $I=1/2$ should occur already at the level of the pentaquark baryons~\cite{jaffe-w}.

A large mixing between $a^+_{c \bar s}$  and $f^+_{c \bar s}$  
leads to decays of the mass eigenstates that do not respect the 
isospin symmetric pattern given above.
To be quantitative, assume that the mass eigenstates are superposition 
of the two, OZI conserving, eigenvectors:
\begin{eqnarray}			
&&|S_u\rangle =[cu] [\bar u \bar s], \nonumber \\
&&|S_d\rangle= [cd] [\bar d \bar s].
\end{eqnarray}
According to:\\
\begin{eqnarray}
&&  |D_h\rangle = \cos\theta |S_u\rangle+\sin\theta 
|S_d\rangle, \nonumber \\ &&|D_l \rangle = -\sin\theta |S_u\rangle+\cos\theta |S_d\rangle.
\end{eqnarray}
Decay amplitudes of four-quark states are computed following Ref.~\cite{us}, 
in terms of a single amplitude $A$.
Keeping into account the antisymmetric structure of the diquarks, one finds 
easily the results in Table~I. 
$X_q$ is the projection on the $\eta$ meson of the isosinglet 
pseudoscalar state $\eta_q$:\\
\begin{eqnarray}
\eta_q &=& \frac{(u \bar u +d \bar d)}{\sqrt{2}}, \nonumber \\X_q &=&\frac{(\cos\phi +\sqrt{2} 
\sin\phi)}{\sqrt{3}}\simeq 0.72.
\end{eqnarray}Where $\phi$ is the $\eta\eta'$ meson mixing 
angle, $\sin\phi\simeq 0.19$ (quadratic mass formulae).

\begin{table}
\begin{center}
\begin{tabular}{|l|l|l|l|l|}
\hline
      & $D_s\eta$    & $D_s\pi^0$ & $D^0K^+$ & $D^+ K^0$\\
\hline
$D_h$      & $\frac{A(\cos\theta+\sin\theta)X_q}{\sqrt{2}}$   & 
$\frac{A(\cos\theta-\sin\theta)}{\sqrt{2}}$ &
$-A \cos\theta$ & $-A\sin\theta$\\
\hline
$D_l$           & $\frac{A(\cos\theta-\sin\theta)X_q}{\sqrt{2}}$               
& $-\frac{A(\cos\theta+\sin\theta)}{\sqrt{2}}$ &
$A\sin\theta$ & $-A\cos\theta$\\
\hline
\end{tabular}
\end{center}
\caption{ Amplitudes for the decays of $D_h$ and $D_l$ in 
the OZI allowed channels.}
\end{table}
The particle produced in strong reactions is a mixture of the two eigenstates 
according to the respective probabilities: $P=$~prob. of producing $D_h$, 
$(1-P)=$~prob. of producing $D_l$. Apart from a normalization factor:
\begin{equation}P=|A^{(0)}(\cos\theta + \sin\theta)/\sqrt{2}+ 
A^{(1)}(\cos\theta - \sin\theta)/\sqrt{2}|^2
\end{equation}
and $A^{(0,1)}$ are the amplitudes to produce the isospin 1 and 0 states, 
$a$ and $f$. The decay probability of the $D_h/D_l$ mixture into a given 
channel, $X$,  is:
\begin{equation}
\Gamma (X)= P \Gamma_h(X)+ (1-P)\Gamma_l(X).
\end{equation}
The ratio of the $D^0K^+$ to the $D_s^+\eta$ rates is computed 
assuming S-wave decays. Using the 
amplitudes of Table~I we find:
\begin{eqnarray}
&&R^0 = \frac{\Gamma(D^0K^+)}{\Gamma(D_s^+\eta)}
\frac{X^2_q p_{D_s\eta}}{2 p_{D^0 K^+}} 
\approx 0.027  \nonumber\\
&&=\frac{P\cos^2 \theta  + (1-P)\sin^2 \theta}{(1+\sin 2\theta)P 
+ (1- \sin 2\theta)(1 -P)}.
\label{eq:err}
\end{eqnarray} 
\newline
We give in Fig.~1 the curve representing~(\ref{eq:err}) 
in the $\theta - P$ plane. 
\begin{figure}[ht]
\begin{center}
\epsfig{
height=6.5truecm, width=8truecm,
        figure=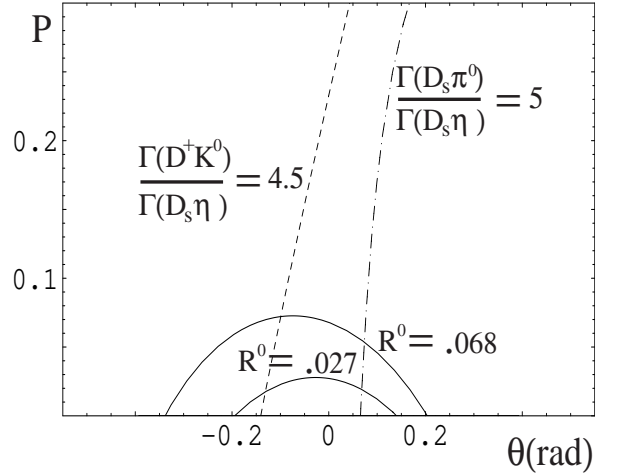}
\caption{\label{fig.1} \footnotesize
Solid lines: $P$ versus $\sin \theta$ according 
to Eq.~(\ref{eq:err}) for $R^0=0.027$ (see text) and
$0.068$. Dotted and dot-dashed lines 
are the corresponding curves for $D_s^+ K^0$ and 
$D^+\pi^0$ modes. Bounds on the rates given in the text
normalized to the $D_s \eta $ rate are found by 
requiring these curves to intersect the solid line 
with $R^0=0.027$.
}
\end{center}
\end{figure}
The very small value of $R^0$ reflects into an allowed region 
with very small $P$ and $\theta$. We find:
\begin{equation}
-0. 19 <\sin \theta < + 0.14,\,\,\,\,  P< 0.03.
\end{equation}
The picture that emerges is that $D_{sJ}(2632)$ is to 
high precision $D_l$, which in turn is mainly $S_d$, whose decay 
into $D^0K^+$ is OZI forbidden with only a small component along $S_u$ 
for which $D^0 K^+$ is OZI allowed.
We report in the same figure 
two similar curves referring to the $D^0K^+$ and $D_s\pi^0$ 
modes, computed for the indicated value of the ratio of the rates 
to the $D_s\eta$ mode. These curves intersect the first one for 
values in the intervals:
\begin{equation}
4 < \frac{\Gamma(D^+ K^0)}{\Gamma(D_s\eta)} < 7.6 ; \,\,\,\,
1.7 < \frac{\Gamma (D_s \pi^0)}{\Gamma(D_s\eta)} < 6.5.
\end{equation}
A last comment refers to the doubly charged, exotic
state: $a^{++}_{c\bar s}=[cu] [\bar d \bar s]$, 
expected to decay into $D_s\pi^+$ or $D^+K^+$. 
The smallness of $P$ indicates an almost complete cancellation: 
$A(0)+A(1)\simeq 0$. 
However, the state $a^{++}_{c \bar s}$  is produced with the 
amplitude $A(1)$ only and is thus expected 
to be produced about as much as the $D_{sJ}(2632)$

{\sl Acknowledgments}. 
ADP thanks the Physics Department of the
University of Bari for their kind hospitality.

\end{document}